# Mu2e upgrade physics reach optimization studies for the PIP-II era

**V. Pronskikh**[1]
*Fermi National Accelerator Laboratory, USA*
E-mail: `vspron@fnal.gov`

**D. Glenzinski**
*Fermi National Accelerator Laboratory, USA*
E-mail: `douglasg@fnal.gov`

**N. Mokhov**
*Fermi National Accelerator Laboratory, USA*
E-mail: `mokhov@fnal.gov`

**R. Tschirhart**
*Fermi National Accelerator Laboratory, USA*
E-mail: `tsch@fnal.gov`

The Mu2e experiment at Fermilab is being designed to study the coherent neutrino-less conversion of a negative muon into an electron in the field of a nucleus. This process has an extremely low probability in the Standard Model and its observation would provide unambiguous evidence for BSM physics. The Mu2e design aims to reach a single-event-sensitivity of about 2.5 x $10^{-17}$ and will probe effective new physics mass scales in the $10^3$-$10^4$ TeV range, well beyond the reach of the LHC. This work examines the maximum beam power that can be tolerated for beam energies in the 0.5-8 GeV range exploring variations in the geometry in the region of the production target using the MARS15 code. This has implications for how the sensitivity might be further improved with a second generation experiment using an upgraded proton beam from the PIP-II project, which will be capable of providing MW beams to Fermilab experiments later in the next decade.

[1]Speaker

## 1. Introduction

The Mu2e experiment at Fermilab [1] will search for evidence of charged lepton flavor violation (cLFV) by observing the conversion of a negative muon into an electron in the Coulomb field of a nucleus without emission of neutrinos. This process is extremely suppressed in the Standard Model, but is predicted to occur at rates observable by Mu2e in a wide variety of new physics models. The Mu2e experiment will probe effective new-physics mass scales in the $10^3$-$10^4$ TeV energy range. One of the main parts of the Mu2e experimental setup is its target station in which negative pions are generated in interactions of the 8-GeV primary proton beam with a tungsten target, which will be capable of producing $\sim 2 \cdot 10^{17}$ negative muons per year. A large-aperture 5-T superconducting production solenoid (PS) enhances pion collection, and an S-shaped transport solenoid (TS) delivers muons and pions to the Mu2e detector. A heat and radiation shield (HRS), installed on the inner bore of the PS, mitigates the effects of radiation dose and heat deposition to protect the PS and the first TS coils from damage. The muons traversing the TS are stopped on an aluminum stopping target situated in the upstream portion of a large aperture detector solenoid (DS), which also houses a tracker and calorimeter in the downstream portion that precisely determine the momenta and energy of particles originating in the stopping target. The Mu2e experiment has a design sensitivity $10^4$ times better than previous muon-to-electron conversion experiments and is scheduled to begin commissioning in 2020 [2].

## 2. Simulations

Regardless of the Mu2e outcome, a next generation experiment, Mu2e-II, with a sensitivity extended another factor of 10 or more, offers a compelling physics case [3]. The improved sensitivity would be enabled by the proposed PIP-II upgrade project, which would significantly improve the Fermilab proton source to enable next-generation intensity frontier experiments [4].

PIP-II is a proposed 250-meter long linac capable of accelerating a 2-mA proton beam to a kinetic energy of 800-MeV corresponding to 1.6-MW of power. Most of the beam will be utilized for the Fermilab Short Baseline Neutrino and Long Baseline Neutrino Facility neutrino programs, but about 200-kW of 800-MeV protons will be available for additional experiments. To achieve another factor of ten improvement in sensitivity, Mu2e-II will require about 100-kW. The linac will have the possibility of being further upgraded to proton energies as high as 3-GeV.

The present Mu2e design is optimized for 8-kW of protons at 8 GeV. This work uses a MARS15 simulation [5] to study the radiation damage to the PS coils, the peak power deposition in the PS coils, and the stopped muon yield as a function of proton beam energy 0.5 – 8 GeV range. The radiation damage is quantified as displacements-per-atom (DPA). An optimal beam energy would maximize the stopped muon yield while minimizing the DPA damage and heat deposition in the PS coils. For these studies the current Mu2e geometry and beamline are assumed to work for all proton energies so that the simulations begin with the protons interacting on the tungsten production target. This assumption is the focus of a separate study.

The MARS15 Mu2e model is shown in Figure 1. The model includes a description of the superconducting coils in the PS, TS, and DS and the surrounding cryostats as well as collimators situated along the TS for momentum and sign selection. The HRS is situated along the length of the PS inside the cryostat. The tungsten production target is 0.6-cm in diameter and 16-cm long radially centered in the HRS. The aluminum stopping target is modeled as 17 aluminum foils, each 200-μm thick with a ~10 cm diameter, placed 5 cm apart radially, centered in the upstream portion of the DS. For these studies the HRS material was assumed to be either bronze or tungsten. All relevant processes were simulated including production of pions of both signs, their transport, their decay into muons, muon transport and decay, and the stopping of muons in the aluminum target. In the MARS15 simulations, the LAQGSM [6] generator was used for pion and neutron production as well as for other high-energy particle interactions. The MCNP [7] model based on ENDFB-VI [8] was used for neutron transport below 14-MeV.

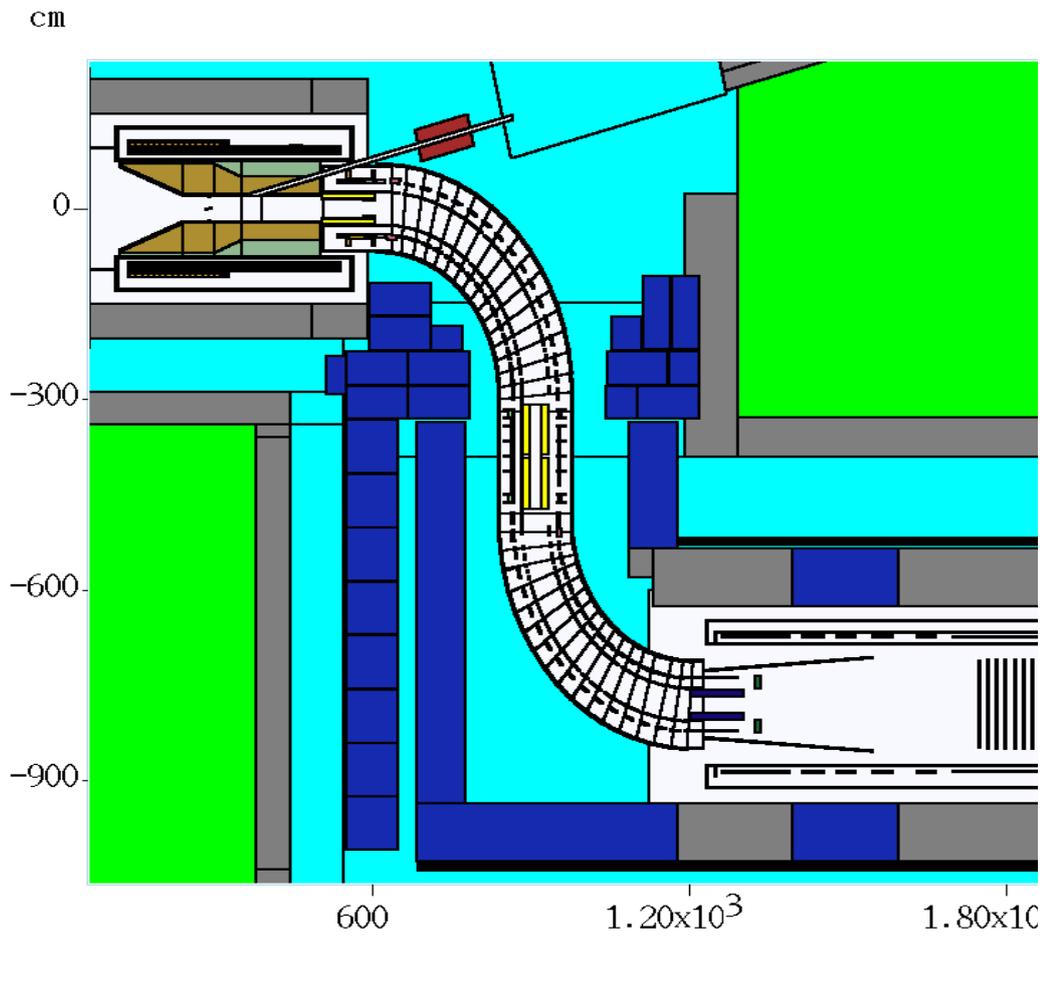

Figure 1. The MARS15 model of Mu2e used in these studies. The PS is located in the top left corner. The incoming proton beam arrives diagonally from right. The S-shaped TS solenoid includes a collimator at each end and at its center for momentum and charge selection. The DS is located in the lower right corner. The aluminum stopping target foils are not shown. At the right end of the DS the first several tracker stations are shown. The blue and grey blocks represent concrete shielding blocks.

To simulate DPA in the magnet coils the NRT [9] model implementation in MARS15 was used. Below 150 (20) MeV the NRT it uses the FermiDPA 1.0 [10] cross-section library, which is based on NJOY [11] (ENDFB-VII). The distributions indicate that the neutron-induced component contributes up to 70% of the total DPA (the rest is photon- and proton-induced).

The limit on DPA was determined based on the requirement that the residual resistivity ratio (RRR), with an initial value of 600, remains at a value of 100 or larger. The calculations involved experimental uncertainties (blue and green lines) and used the KUR reactor data measured by the COMET collaboration [12]. Figure 4 indicates that the peak DPA damage to the coils should not exceed the $\sim 5 \cdot 10^{-5}$ before annealing. Experiments show that room temperature annealing of Al-stabilized NbTi can fully recover the RRR. Mu2e plans to anneal the PS coils once a year.

The number of stopped muons was estimated by convoluting the momentum spectra of muons and pions arriving at the TS entrance with an acceptance function corresponding to the probability of a muon stop as a function of particle momentum. A G4Beamline simulation [13], which provides a fast and accurate beam-transport model, was used to derive acceptance functions separately for muons and pions.

3. Results

Radiation damage to the superconducting coils of the Mu2e production solenoid (quantified as displacements per atom, DPA), the power deposition limitations on the performance of the Mu2e production solenoid, and the muon-stopping rate are calculated for a variety of potential beam-upgrade scenarios that would enable a next generation Mu2e-II experiment. Utilizing a MARS15 simulation model that includes a detailed description of the current Mu2e beamline, shielding, and detector geometry the above quantities are estimated for proton beam energies in the range 0.5 – 8 GeV for beam intensities up to 100 kW. A heat and radiation shield (HRS) located inside the production solenoid mitigates the radiation damage and power deposition effects on the superconducting coils. In these studies the HRS was assumed to be composed of either bronze or tungsten. The DPA, power deposition and muon-stopping rate all peak at a proton beam energy of 2-3 GeV.

Taking at face value the current estimates of the maximum DPA the PS coils can tolerate before performance is affected, these MARS15 simulations suggest that the current production solenoid with its HRS could tolerate beam power of <100 kW in the 0.8 – 8 GeV range of beam energies. An ad hoc figure-of-merit is constructed from the ratio of stopped muons to DPA in the production solenoid coils and suggests that the optimal proton beam energy would be in the 1-3 GeV range. The figure-of-merit is up to about 30% worse outside of this optimal range over the range of energies considered. It is about the same at a beam energy of 0.8 GeV as it is at 8 GeV.

The stopped muon yields are combined with the peak DPA estimates discussed above to obtain the figure-of-merit as a function of proton energy. By this metric the optimal beam energy lay in the 1-3 GeV range and varies slowly above about 4 GeV. The figure-of-merit is about the same at 800 MeV as it is at 8 GeV proton beam energy.

The estimate of the single-event-sensitivity can be repeated assuming instead that the DPA limitations discussed above cannot be mitigated and thus limit the maximum beam power that can be tolerated at a given proton beam energy. This will reduce the total number of stopped muons that can be produced for a given run time and thus degrade sensitivity. The resulting estimates, for a nominal 3 year run at the maximum beam power allowed assuming DPA limitations, as a function of proton beam energy are shown in Figure 2. This may be considered a pessimistic scenario.

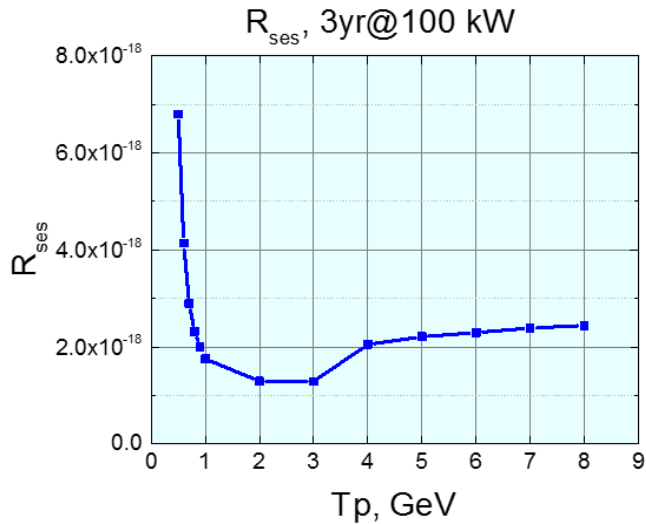

Figure 2. An estimate of the single-event-sensitivity (SES) for a nominal 3-year run at 100 kW for various proton beam energies ($T_p$). The assumptions affecting this estimate are discussed in the text. By comparing to the estimated Mu2e SES ($2.5 \times 10^{-17}$ for 3 years of 8 kW of proton beam at 8 GeV), this suggests that a next-generation Mu2e-II experiment might plausibly improve the sensitivity by a factor of 10 utilizing an upgraded proton source. The assumptions here might be considered optimistic.

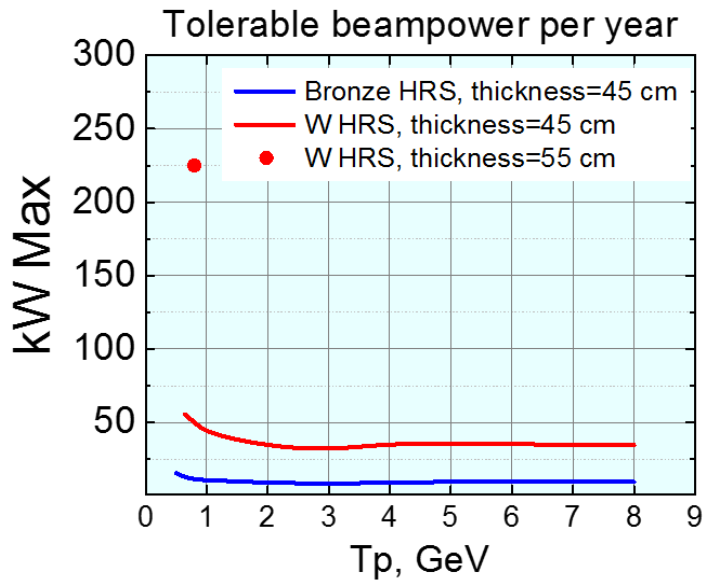

Figure 3. An estimate of the maximum beam power per year that can be tolerated by the current HRS design (both bronze and tungsten versions). The red dot corresponds to the tungsten HRS with an increased thickness of 55 cm (due to decreased inner bore) as estimated by extrapolation

of the dependence of DPA damage on material thickness at the projected PIP-II energy of 800 MeV.

A single-event-sensitivity is estimated for Mu2e-II, as described in the text, as a function of proton beam energy assuming the beam power is limited by the DPA damage to the PS coils and its detrimental effects to the superconductor residual resistivity ratio. More optimistically, if the DPA effects can be mitigated to allow running at 100 kW, the single-event-sensitivity can improve by a factor of 10 or more for a nominal 3 year run utilizing an aluminum stopping target (cf. Figure 2). An optimistic estimate of the HRS inner bore decrease (cf. Figure 3) based on extrapolation suggests that the tungsten HRS with a 55 cm thickness (which is likely safe for the muon yield) will likely tolerate a beam power of more than 50 kW per year.

Strategies to extend the DPA limitations of the production solenoid are under study and include, for example, the possibility of more frequent anneals, an improved HRS geometry, or an improved production solenoid. Other parts of the Mu2e apparatus may also have limitations that would imply a beam intensity of < 100 kW. These are the subject of separate on-going studies.